\documentclass[aoas,MSNbibl,nameyear,seceqn,dvips]{arximspdf}
\usepackage{dcolumn}
\usepackage{graphicx}

\doi{10.1214/12-AOAS537} 
\volume{6}
\issue{3}
\pubyear{2012}
\firstpage{928}
\lastpage{949}

\makeatletter
\newcolumntype{d}[1]{D{.}{.}{#1}}

\setattribute{abstract}{width}{281pt}

\makeatother

\begin{document}
\begin{frontmatter}

\title{A model for sequential evolution of ligands by exponential enrichment (SELEX) data}
\runtitle{A model for SELEX}

\begin{aug}
\author[a]{\fnms{Juli}~\snm{Atherton}\corref{}\thanksref{t1,t3,t4}\ead[label=e1]{JuliAtherton@gmail.com}},
\author[b]{\fnms{Nathan}~\snm{Boley}\thanksref{t1,t5}\ead[label=e2]{npboley@gmail.com}},
\author[b]{\fnms{Ben} \snm{Brown}\thanksref{t5}\ead[label=e3]{ben@newton.berkeley.com}},
\author[c]{\fnms{Nobuo}~\snm{Ogawa}\thanksref{t5}\ead[label=e4]{nobogw@gmail.com}},
\author[c]{\fnms{Stuart M.}~\snm{Davidson}\thanksref{t5,t6}\ead[label=e5]{stuartd@horizoncable.com}},
\author[c]{\fnms{Michael B.}~\snm{Eisen}\thanksref{t5}\ead[label=e6]{mbeisen@gmail.com}},
\author[c]{\fnms{Mark~D.}~\snm{Biggin}\thanksref{t5}\ead[label=e7]{mdbiggin@lbl.gov}}
\and
\author[b]{\fnms{Peter}~\snm{Bickel}\thanksref{t3}\ead[label=e8]{bickel@stat.berkeley.edu}\ead[label=u1,url]{http://bdtnp.lbl.gov/Fly-Net/}}

\thankstext{t1}{Joint first authorship.}
\thankstext{t3}{Supported in part by NIH-R01GM075312.}
\thankstext{t4}{Supported in part by NSERC Grant RGPIN 356107-2009.}
\thankstext{t5}{The in vitro and in vivo DNA binding data were funded
by the U.S. National Institutes of Health (NIH) under Grant GM704403 (to MDB and MBE).
Work at Lawrence Berkeley National Laboratory was conducted
under Department of Energy Contract DE-AC02-05CH11231.}
\thankstext{t6}{Not affiliated with Lawrence Berkeley National Laboratory now.}
\runauthor{J. Atherton et al.}

\affiliation{Universit\'e du Qu\'ebec \`a Montr\'eal,
University of California, Berkeley,
University of California, Berkeley,
Lawrence Berkeley National Laboratory,
Lawrence Berkeley National Laboratory,
Lawrence Berkeley National Laboratory,
Lawrence Berkeley National Laboratory
and
University of~California, Berkeley}
\address[a]{J. Atherton\\
D\'epartement de Math\'ematiques \\
Universit\'e du Qu\'ebec \`a Montr\'eal (UQAM)\\
Canada\\
\printead{e1}}

\address[b]{N. Boley\\
B. Brown\\
P. Bickel\\
Department of Statistics\\
University of California, Berkeley\\
USA\\
\printead{e2}\\
\phantom{E-mail:} \printead*{e3}\\
\phantom{E-mail:} \printead*{e8}}

\address[c]{N. Ogawa\\
S. Davidson\\
M. Eisen\\
M. Biggin\\
Genomics Division\\
Lawrence Berkeley National Laboratory\\
USA\\
\printead{e4}\\
\phantom{E-mail:} \printead*{e5}\\
\phantom{E-mail:} \printead*{e6}\\
\phantom{E-mail:} \printead*{e7}\\
\printead{u1}}
\end{aug}

\received{\smonth{4} \syear{2010}}
\revised{\smonth{11} \syear{2011}}

%
\begin{abstract}
A Systematic Evolution of Ligands by EXponential enrichment (SELEX) experiment
begins in round one with a random pool of oli\-gonucleotides
in equilibrium solution with a target. Over a few
rounds, oligonucleotides having a high affinity for the target are selected.
Data from a high throughput SELEX experiment consists of lists of
thousands of
oligonucleotides sampled after each round. Thus far, SELEX experiments
have been
very good at suggesting the highest affinity oligonucleotide, but
modeling lower affinity recognition site variants
has been difficult. Furthermore, an alignment step has always been
used prior to analyzing SELEX data.

We present a novel model, based on a biochemical parametrization of
SELEX, which
allows us to use data from all rounds to estimate the affinities of the
oligonucleotides. Most notably, our model also aligns the oligonucleotides.
We use our model to analyze a SELEX experiment containing double
stranded DNA
oligonucleotides and the transcription factor Bicoid as the target.
Our SELEX model outperformed other published methods for predicting
putative binding sites for Bicoid as indicated by the results of an
in-vivo ChIP-chip experiment.
\end{abstract}

%

\end{frontmatter}

\section{Introduction}

Transcription factors are proteins that regulate gene transcription of
DNA by binding to DNA sequence motifs within the genome.
Mapping these DNA recognition sequences and determining the
relationship between DNA sequence and transcription factor binding
affinity is central to understanding the regulation of gene expression.
Transcription factors comprise approximately 8\% of the genes encoded
in the human genome. A~comprehensive understanding of the behavior of
these proteins will aid in our understanding of key developmental
processes, including body patterning, brain development and tissue
specification.

One in-vitroassay, known as Systematic Evolution of Ligands by
EXponential enrichment (SELEX), indirectly measures the affinity of a
transcription factor binding to various DNA sequences. SELEX was
introduced in the 1990s by \citet{Gold} and \citet{Szostak}. It has been
used in a number of genomic studies [e.g., \citet{SR} and \citet{CopR}]
and for the purposes of drug discovery [e.g., \citet{Cell} and \citet
{Macugen}]. In genomic studies, SELEX has been used to identify the
highest affinity recognition sequences for target proteins.

More recently there has been an emphasis on using SELEX data to
estimate not just the highest affinity sequences but also a matrix for
the free energy of binding. Using the free energy matrix, one can build
a model which takes as input a nucleotide sequence and outputs the
affinity of the sequence for the transcription factor. With a flexible
model, one can scan the genome to find high to medium affinity putative
binding sites. Having such a model is important since the nucleotide
sequence with the highest affinity for the transcription factor might
not be occupied in-vivo. For instance, due to DNA folding and histone
interference, the highest affinity site may be inaccessible to the
transcription factor. Also, the specificity of the site may play a
role. That is, a medium affinity site surrounded by very low affinity
sequences might be a functionally more important binding site than a
high affinity site surrounded by other high affinity sites. Such
requirements have led researchers to consider thermodynamic models for
SELEX. \citet{Djordjevic2006} and \citet{StormoZ} are two thermodynamic
models for SELEX that precede ours. We will clearly illustrate how our
model diverges from \citet{Djordjevic2006} and \citet{StormoZ} in
Sections~\ref{secSELEX} and~\ref{secPS} after we describe the SELEX
experiment in detail.

Our model is a result of a large collaboration, the Berkeley Drosophila
Transcription Network Project (BDTNP). The goal of the BDTNP is to
understand the early developmental transcription factors in fly
embryos. As part of this collaboration, in-vitro SELEX, in-vivo
ChIP-chip and, most recently, in-vivo ChIP-seq have been performed on
many transcription factors. Although the in-vivo ChIP-seq results are
extremely important because they identify regions along the genome to
which a transcription factor actually bound at an instant in time in a
particular developmental stage in a specific tissue or cell lineage, we
believe that the in-vitro SELEX experiment is still extremely relevant
for two reasons.

First, the ChIP-seq assay is exceedingly expensive, currently at a
minimum cost of $\$5\mathrm{K}$ per sample. To fully understand a developmental
process, it would be necessary to conduct ChIP-seq in every tissue or
cell lineage in an animal throughout its development. At $\$5\mathrm{K}$ per
sample this cost is already prohibitive before we even account for the
manpower required. The in-vitro binding data from SELEX allows us to
reason about all locations in a genome that might be bound by the
transcription factor of interest in {\textit{any}} sample. Obviously, this
can be very powerful when combined with information from other in-vivo
assays, such as DNasel accessibility experiments [\citet{Xiao} and \citet
{Kaplan}].

Second, there is great value to obtaining the qualitative,
thermodynamic estimates of protein/DNA binding affinities that we model
in this paper from SELEX data. Ultimately, biologists would like to
understand the relationship between transcription factor binding
patterns and gene expression [\citet{Ahmet}]. Transcription factors have
been shown to work together in complex spatial arrangements in order to
modulate gene expression [\citet{Biggin}] and the dynamics of these
spatial configurations and their effects on transcription initiation
can not be observed by ChIP-seq or any other widely utilized assay.
Such critical aspects of gene regulation can, at present, on a~large
scale, be inputted only computationally, using models of protein/DNA
binding affinities [\citet{Ravasi}, \citet{Alan} and \citet{Kaplan}].
Therefore, models such as the one we propose here based on SELEX data
will continue to be an important area of computational biology for the
foreseeable future.

\section{The SELEX assay and likelihood for the model}\label{secSELEX}
A typical SELEX experiment begins in round one with a solution of
random double stranded DNA oligonucleotides and a transcription factor.
In the application presented in this paper the oligonucleotides are 16
base pairs long sequences and are flanked by additional DNA sequences.

\begin{figure}

\includegraphics{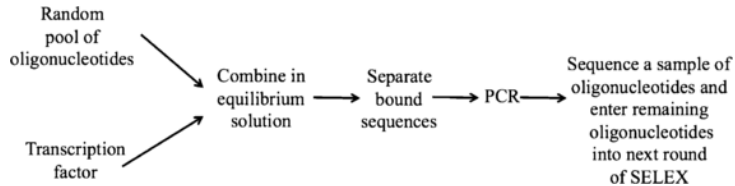}

\caption{The main experimental steps for round one of a SELEX experiment.}
\label{figexample}
\end{figure}

The oligonucleotides react with the transcription factor and eventually
a~dynamic equilibrium is reached where the concentrations of bound
oligonucleotides, unbound oligonucleotides and unbound target are constant.
After equilibrium is reached, the bound oligonucleotides are separated
from the solution.
Next, a polymerase chain reaction (PCR) is performed on the
oligonucleotides sampled from the end of round one.
PCR chemically amplifies the quantity of DNA present in a way that does
not significantly change the frequency distribution of oligonucleotides.
At this point, a sample is taken for sequencing, and the remaining
oligonucleotides are entered into round two.
The main steps for round one of SELEX are depicted in Figure~\ref{figexample}.

Round two of SELEX proceeds exactly as round one, except that the
initial pool of oligonucleotides is the set of bound oligonucleotides
from round one that went through PCR but were not sequenced.
Thereafter, the assay proceeds as before: the oligonucleotides react
with the transcription factor and, after equilibrium is reached, the
bound oligonucleotides are selected and PCR is performed.
A sample is taken for sequencing and the remaining oligonucleotides are
entered into round three.
These steps are repeated for as many rounds as the experimenter
desires; see \citet{Ogawa} for full experimental details.

The outcome of a SELEX experiment is observed by sequencing the
oligonucleotides that are sampled at the end of each round.
That is, after performing the assay, the results are a list of
sequenced oligonucleotides and usually meta-data, such as the SELEX
round in which each oligonucleotide was sequenced, the concentration of
unbound transcription factor in a particular round, and/or the
temperature at which the experiment was performed.

Each sequence is denoted by $S_i$, where $i$ enumerates over all the
different sequence types. Letting $k$ represent the length of the
sequences and carefully accounting for palindromes and reverse
palindromes, for our double stranded DNA application
$i= 1, \ldots,n$ where $n= 2^{k-1} + 4^{k-1}$.
We let $r$ identify the round number beginning with $r=0$ for the
initial random pool of sequences. The number of times sequence type
$S_i$ is observed in round $r$ is represented by $l_{i,r}$. Table \ref
{tabSeq} shows the first ten 16mers $S_i$ and the number of times each
sequence appeared $l_{ir}$ in round $r=3$ of a SELEX experiment for the
transcription factor Bicoid. Although we only show ten sequences, a
total of $1324$ unique sequences $S_i$ were observed in round 3 of the
SELEX experiment depicted in Table~\ref{tabSeq}.

\begin{table}
\tablewidth=220pt
\caption{Example of first ten sequences $S_i$ and their frequencies
$l_{i,3}$ collected after the third round of a~SELEX experiment for the
transcription factor Bicoid}\label{tabSeq}
\begin{tabular*}{220pt}{@{\extracolsep{\fill}}lc@{}}
\hline
$\bolds{S_i}$ & $\bolds{l_{i,3}}$\\
\hline
{\texttt{TCCCATTAATCCCACC}} & 2 \\
{\texttt{GGTGTCGGTTTAAGCG}} & 2\\
{\texttt{CTGATTAATCCGAGTG}} & 1\\
{\texttt{TGAGATTCCATACCCT}} & 1\\
{\texttt{TGTGAGGATATGTTTC}} & 1\\
{\texttt{TGGGGTTGGATTAAAG}} & 1\\
{\texttt{GGATTAGGGTTAAGCA}} & 1\\
{\texttt{GACCCCGGCCTAATCC}} & 1\\
{\texttt{GGTAATCTCGGGATTA}} & 1\\
{\texttt{TGGACGGATTACGCGG}} & 1\\
\hline
\end{tabular*}
\vspace*{-2pt}
\end{table}

A complicating factor of SELEX is that the length of the binding
site~$l$ to the transcription factor is less than the length of the
sequences $k$. In the application of this paper we have $k=16$ and we
estimate the binding site length of Bicoid $l$ to be at most $10$. All
previous methods, including \citet{Djordjevic2006} and \citet{StormoZ},
for analyzing SELEX data use an alignment step prior to analyzing the
SELEX data. Such aligners [e.g., Multiple Em for Motif Elicitation
(MEME), \citet{MEME}] are not based on the thermodynamics of binding.
For each kmer, these aligners will output their ``best guess'' for the
lmer to which the sequence $S_i$ is bound. We denote the lmer binding
sites by $b_j$. For example, Table~\ref{tabSeqAlign} shows ten aligned
sequences from a SELEX experiment for the transcription factor Bicoid.
The sequences are aligned for a binding site of $l=8$ using an aligner
written in the Biggin lab by Stuart Davidson.

\begin{table}[b]
\tablewidth=220pt
\caption{Ten aligned sequences from a SELEX experiment for the
transcription factor Bicoid. The sequences here were aligned assuming a
binding site of length $l=8$}\label{tabSeqAlign}
\begin{tabular*}{220pt}{@{\extracolsep{\fill}}lcc@{}}
\hline
& $\bolds{b_j}$ & \\
 \hline
{\texttt{ATA}} & {\texttt{TTAATCCG}} & {\texttt{ATAAC}} \\
{\texttt{CACCC}} & {\texttt{TAAATCTT}} & {\texttt{CGT}} \\
& {\texttt{TTAATCCA}} & {\texttt{GCGCATCA}} \\
{\texttt{ACCC}} & {\texttt{TTAATCCC}} & {\texttt{CCCA}} \\
{\texttt{CAACC}} & {\texttt{TTAATCCC}} & \\
{\texttt{TAA}} & {\texttt{TCCCTCCT}} & {\texttt{AATCC}} \\
{\texttt{T}} & {\texttt{TTAATCCT}} & {\texttt{GATCCCC}} \\
{\texttt{GGA}} & {\texttt{TTAACTCG}}&{\texttt{GATTA}} \\
{\texttt{GAGAGG}} & {\texttt{TTAATCCA}} & {\texttt{CT}} \\
{\texttt{GTAC}} &{\texttt{CAAGTCAC}} &{\texttt{CACA}}\\
\hline
\end{tabular*}
\end{table}

Previous models for SELEX take the estimated binding site sequences
$b_j$ from an aligner as input. Our model selects binding sites
dynamically as part of the optimization. That is, the model takes the
full kmer $S_i$ sequences that were sequenced after each round of the
SELEX experiment as input. The likelihood (\ref{eqnlhd}) is
parametrized in terms of $P_r(S_i)$, where $P_r(S_i)$ denotes the
probability of selecting sequence $S_i$ in round $r$. In Section \ref
{secPS} we provide the parametrization for $P_r(S_i)$ in terms of the
free energy, $\Delta G$, a thermodynamic measure of affinity. Letting
$R$ denote the total number of rounds for the SELEX experiment, we have

\begin{equation}\label{eqnlhd}
L (\Delta G|l_{11}, \ldots, l_{nR}) = \prod_{r = 1}^R
\Biggl( \prod_{i = 1}^n P_r (S_i)^{l_{ir}} \Biggr).
\end{equation}

It is easily seen from the likelihood (\ref{eqnlhd}) that our model
for SELEX can take as input data from all rounds of a SELEX experiment.
This is important, as there is evidence that a range of affinities is
required to properly estimate the free energy, $\Delta G$ [see the
review article \citet{Djordjevic2007}]. Our model for SELEX is the first
model to use data from all rounds of the experiment; previous models
use only data from the last round which consists of high affinity sequences.

\section{Parametrization of the model}\label{secPS}
Section~\ref{secprob} describes how the probability of
a sequence $S_i$ binding to the transcription factor in round $r$,
$t_r(S_i)$, is parametrized in terms of the Gibbs free energy $\Delta
G$. Section~\ref{secPSsub} provides the parametrization of the
probabilities $P_r(S_i)$ of drawing $S_i$ from round $r$. Appendix~\ref{secchem}
gives the necessary chemical background.

\subsection{Probability of a sequence $S_i$ binding}\label{secprob}
In SELEX, we have multiple oligonucleotide types $S_i$ in solution. At dynamic
equilibrium in round $r$, the probability of any copy of type $S_i$
being bound at a particular
instant is equal to the fraction of $S_i$, that is, bound, $t_r(S_i)$.
Letting $[\mathit{TF}\dvtx S_i]_r$ and $[S_i]_r$ represent the long term average
concentrations of the bound product and unbound sequences $S_i$ in
round $r$, we have
%
\begin{equation}\label{eqnfrac}
t_r(S_i) = \frac{[\mathit{TF}\dvtx S_i]_r }{[\mathit{TF}\dvtx S_i]_r +
[S_i]_r }.
\end{equation}

We are interested in modeling the affinity of oligonucleotides that
bind in a sequence specific manner to the target. Specific binding
involves hydrogen bonding, van der Waals interactions and other
short-range forces.
Sequence independent binding also occurs. This is due in part because
oligonucleotides bind weakly via electrostatic forces [see {\citet
{vonHipple}}], and because a~small percentage of DNA will
nonspecifically associate with the bead or non-DNA binding surfaces of
the target.
Thus, even oligonucleotides that do not bind to the target specifically
can be present in later rounds. We make three assumptions concerning
specific binding for any oligonucleotide type~$S_i$:

\begin{longlist}[1.]
\item[1.] All identical copies of the same oligonucleotide type $S_i$ bind
at the same subsequence $b_j$. We refer to this subsequence as the
binding site.\vadjust{\goodbreak}
\item[2.] The subsequence $b_j$ is assumed to be of fixed length $l$ and
independent of the oligonucleotide type $S_i$ in which it is contained.
\item[3.] The binding site $b_j$ for each oligonucleotide type $S_i$ is
that subsequence which has maximum affinity according to the proposed model.
\end{longlist}
These correspond to the assumptions that the binding affinity
of the sequence
is solely a function of the binding site and that there is only one binding
site per oligonucleotide.

Given these assumptions and letting $[\mathit{TF}]_r$ represent the long term
average concentration of unbound transcription factor at dynamic
equilibrium in round $r$, we can use (\ref{eqnrelation}), (\ref
{eqnK}) and (\ref{eqnfrac}) to write
%
\begin{equation}\label{eqnprob}
t_r(S_i) = \frac{[\mathit{TF}]_r \exp({(- \Delta G(S_i))}/{(R_{\mathrm{Gas}}T)})}{1+ [\mathit{TF}]_r \exp({(- \Delta G(S_i))}/{(R_{\mathrm{Gas}}T)})},
\end{equation}
where $\Delta G(S_i) \equiv\Delta G(b(S_i))$ and $b(S_i)$ maximizes
$\Delta G$ among all $b_j$'s
of the length~$l$ we have specified contained in $S_i$.

In a SELEX experiment, $t_r(S_i)$ can also be viewed as the
conditional probability that a particular molecule of the species $S_i$
is bound
at the end of round $r$ given that it is present at the beginning of
round $r$.
Formally,
%
\begin{equation}\label{eqntheoryBoundProb}
t_r (S_i) \equiv P [S_i  \mbox{ bound at the end of }  r    |
\mbox{it is present in }  r].
\end{equation}

Defining $\widehat{[\mathit{TF}]}_r$ to be the concentration of the
transcription factor at a~particular instant, we obtain $\widehat{t_r}(S_i)$,
%
\begin{equation} \label{eqntau}
\widehat{t_r}(S_i) = \frac{\widehat{[\mathit{TF}]_r} \exp({(-
\Delta G(S_i))}/{(RT)})}{1+ \widehat{[\mathit{TF}]_r} \exp({(- \Delta G(S_i))}/{(RT)})},
\end{equation}
which is an estimate of $t_r(S_i)$. We expect that the instantaneous
concentrations $\widehat{[\mathit{TF}]}_r$, $\widehat{[\mathit{TF}\dvtx S]}_r$ and $\widehat
{[S]}_r$ will vary within $5\%$ of their long term average
concentrations $[\mathit{TF}]$, $[\mathit{TF}\dvtx S]$ and $[S]$.

It is very difficult to measure the amount of transcription factor that
is ``active'' in a binding reaction, versus denatured or otherwise
nonfunctional. Hence, we do not have measurements for $\widehat
{[\mathit{TF}]_r}$ and this causes an identifiability problem when estimating
$\Delta G$. The problem is easily remedied by estimating $\Delta\Delta
G$ instead. See Appendix~\ref{appdeltadeltaG} for further discussion.

Although the notation differs, our formulation for the probability of
a~sequence type $S_i$ binding in round $r$, $\widehat{t_r}(S_i)$,
resembles the parametrization first introduced by \citet{Djordjevic2006}
and later used by \citet{StormoZ}.
The thermodynamic formulation (\ref{eqnprob}) includes competitive
binding between oligonucleotides $S_i$ since the $S_i$ are all
competing for the unbound transcription factor. As we search all
possible binding sites of each oligonucleotide type $S_i$ for the
optimal site, our model takes alignment into account implicitly, unlike
\citet{Djordjevic2006} and \citet{StormoZ} which either use a
pre-alignment step as in Table~\ref{tabSeqAlign} or work on data with $k=l$.

\subsection{Probability of drawing a sequence $S_i$}\label{secPSsub}
Next we express the distribution of bound sequences in terms of
(\ref{eqntau}).
We first assume
that each sequence is present in an initial amount $C_0$ in round
zero.
We
then make the assumption that each PCR step replicates each molecule of
type $S_i$
$A_r$ times on average in round $r$. Then, after the $r$th round of
selection the amount of $S_i$ is
\[
C_0 \prod^{\bar{r}}_{r = 1} A_r \widehat{t_r} (S_i).
\]
Dividing the total amount of $S_i$ after round $\bar{r}$ by the total
amount of
all
sequences after round $\bar{r}$ gives an estimate of the frequency
distribution of
bound sequences at the end of round $\bar{r}$. Formally,
%
\begin{equation}\qquad
\label{eqnnojunkprobM} P_{\bar{r}}(S_i) = P[  S_i \mbox{ is sequenced in round } \bar{r}  ] = \frac{\prod_{r = 1}^{\bar{r}} \widehat{t_r}
(S_i)}{\sum_{\mathrm{all\ }S_j}
\prod_{r = 1}^{\bar{r}} \widehat{t_r} (S_j)}.
\end{equation}

\citet{Djordjevic2006} assume that all the sequences they see in the
last round bound the protein and all the sequences they do not see did
not bind. Hence, their likelihood differs significantly from ours. Like
us, \citet{StormoZ} account for the multinomial sampling in~(\ref
{eqnnojunkprobM}). \citet{StormoZ} also account for extra variability
generated during amplification by PCR. Both \citet{StormoZ} and us fail
to correct for
the case in which zero oligonucleotides of a particular species are
bound in
round $r$. The large oligonucleotide counts makes this a
reasonable approximation. For instance, in the data we study in Section
\ref{secInVivo}, each
16mer species had an average of 65,000 copies in round
zero.\looseness=-1

As discussed in Section~\ref{secprob}, it is possible for
oligonucleotides to make it though the selection step via a
variety of mechanisms, including nonsequence mediated, electrostatic
protein--DNA interaction (nonspecific binding),
DNA--DNA interactions or
DNA--apparatus interactions (experimental error). We account for such
sequences in our model, and refer to the effects that result in their
selection collectively as {\textit{Junk Binding}}. If $c_J$ is a
constant between $0$ and $1$, then we can modify our equations
to allow for junk binding as follows:
\[ \label{eqnprobM}
\widehat{t_r}(c_J, S_i) = \bigl( (1 - c_J)\widehat{t_r}(S_i) + c_J
\bigr).
\]
Our parametrization of the junk binding is different from
\citet{Djordjevic2006} and \citet{StormoZ}
who both use only a~thermodynamic parametrization for the nonspecific binding.

\subsection{Binding model}\label{secbinding}
The {\textit{binding model}} is the relationship between the actual DNA
sequence of a binding site $b_j$ and the free energy $\Delta G$.
So far we have formulated our model in complete generality with respect
to the
binding model.
The most widely applied model is an additive one. The additive model
was used in both \citet{Djordjevic2006} and \citet{StormoZ}. Such a model
assumes that each base pair of
DNA makes some contribution to the total binding affinity independent
of all
other base pairs in the binding site. Representing the nucleotide base
pair at position $k$ in $b_j$
as $o_k$, and letting $\varepsilon_t(o_k)$ represent the indicator function
\[
\varepsilon_t(o_k) = \cases{
1, &  \quad $\mbox{if } o_k = t,$\vspace*{2pt}\cr
0, &\quad $\mbox{otherwise},$}
\]
we write the elements of the energy matrix as $\lambda_{kt}$,
%
\begin{equation}\label{eqnbindingModel}
\Delta G (b_j) = \sum_{k = 1}^l \sum_{t \in \{A,C,G,T\}} \lambda_{kt}
\varepsilon_t
(o_k).
\end{equation}
As before, the length $l$ represents the length of the binding site.
The parameters to be estimated are the $\lambda_{kt}$ from the energy matrix.

It is important to note that our additive model (\ref
{eqnbindingModel}) does not correspond to a Position Weight Matrix
(PWM). In a PWM the nucleotide positions are treated independently. In
our notation this means that the probability of sequence $S_i$ binding
to the transcription factor, $t_r(S_i)$, will equal a product of
probabilities where each probability corresponds to a position in the
sequence and the value of each probability is determined by the
nucleotide at the corresponding position. Our model deviates from such
an independence model in two important ways:
\begin{itemize}
\item By assuming that the binding of a sequence is determined by a
smaller binding site, our model permits considerable dependence between
nucleotide positions and sequences well
separated in hamming distance. If we group the sequences by the binding
sites that give minimal free energy, we see that the distribution of
binding probabilities over sequences is a mixture of probability
distributions, each of which, ignoring thermodynamic considerations,
could be characterized by PWM.
\item Even when the sequence and binding site coincide, that is, when
$k$ and $l$ are equal, the probability of a sequence $S_i$ binding
$t_r(S_i)$ is modeled by a~log odds model. Rearranging equation (\ref{eqntau}),
\[
\log\biggl(\frac{t_r(S_i)}{1-t_r(S_i)}\biggr) = \log([\mathit{TF}]_r) - \frac
{\Delta G(S_i)}{RT}.
\]
\end{itemize}

\section{Optimization}\label{secopt}

This section discusses the optimization of our model. In particular,
Section~\ref{secden} explains how we simulate to simplify the
denominator of $P_r(S_i)$ and Section~\ref{secnum} discusses the
numerics of the optimization procedure. There are three identifiability
issues with our model that are easily overcome. The identifiability
issues are presented in Appendix~\ref{appidentifiability}.

\subsection{Denominator of $P_r(S_i)$}\label{secden}
For $k=16$ the number of oligonucleotide types in the initial random
pool is $2^{15}+4^{15}$. It is infeasible to include all
oligonucleotide types in the denominator of (\ref{eqnnojunkprobM}). We
estimate the denominator using Monte Carlo and take a simple random
sample of oligonucleotides by selecting nucleotide base pairs from a
uniform distribution. Our approach differs from \citet{StormoZ} who
discretized the energy distribution in order to simplify the
denominator before numerically optimizing to estimate the free energy
matrix $\Delta G$.

\subsection{Numerical optimization}\label{secnum}
With regards to our model, a point not yet discussed is the difficulty
of maximizing the likelihood. If $[\mathit{TF}]_r$ is ``small,'' then the
denominator in (\ref{eqntau}) can be approximated by one. The
likelihood~(\ref{eqnlhd}) will simplify, and the optimization between
each alignment step becomes a convex optimization problem. However,
since we avoid making simplifications regarding the concentration of a
transcription factor in each round, the optimization is more difficult,
as discussed below.

There are substantial computational and algorithmic difficulties
in fitting the model. Standard optimization techniques are often
ineffective because the likelihood surface is neither convex nor
differentiable. In particular, the lack of continuous derivatives
makes gradient descent methods like Broyden--Fletcher--Goldfarb--Shanno
(BFGS) [\citet{Wright}]
unstable. In addition,
the lack of convexity means that line search methods [\citet{Nelder}]
tend to
become trapped in local maxima. In view of these considerations, we
have had success using downhill simplex methods [\citet{Powell}] from a
large set of
random starting locations. This method is, empirically, stable. The
software tool presented in the supplementary material [\citet{Atherton12}] implements this method. The
simulations and results in Section~\ref{secInVivo} were all produced
using the provided software tool.

\section{Results}\label{secInVivo}
In Section~\ref{secSim} we demonstrate how our model works on
simulated data.
Section~\ref{secStormo} applies our model to Bicoid SELEX data from
the Biggin Lab. We then compare the estimates from our model to
estimates made from an in-vitro
multiplex assay experiment in the Biggin Lab and to estimates made from
the Binding Energy Estimates using the Maximum Likelihood (BEEML) model
of \citet{StormoZ} in Section~\ref{secmultiassay}.

\begin{table}[b]
\caption{The Gibbs free energy matrix estimated from a SELEX
experiment on the transcription factor Bicoid}
\label{tabmatrix}
\begin{tabular*}{\textwidth}{@{\extracolsep{\fill}}ld{2.6}d{2.6}d{3.6}d{3.6}@{}}
\hline
& \multicolumn{1}{c}{{\texttt{\textbf{A}}}} & \multicolumn{1}{c}{{\texttt{\textbf{C}}}} &
\multicolumn{1}{c}{{\texttt{\textbf{G}}}} & \multicolumn{1}{c@{}}{{\texttt{\textbf{T}}}} \\
\hline
\phantom{0}1 & -4.722516 & -5.729347 & 0.000000 & -6.251779\\
\phantom{0}2&-7.447426 & -5.981440 & 0.000000 & -16.853690 \\
\phantom{0}3 &0.000000 & -6.946246 & -15.701235& -8.529272 \\
\phantom{0}4 &-7.746046 & -15.548042& -12.535315 & 0.000000 \\
\phantom{0}5 &-7.989755 & -7.201358& -24.708969 & 0.000000\\
\phantom{0}6&0.000000 & -9.611195 & -8.497223 & -5.336888\\
\phantom{0}7 &-0.505663 & -19.926999 & 0.000000 & -4.445374\\
\phantom{0}8 &-1.836787 & -0.228140 & 0.000000 & -0.945140\\
\phantom{0}9 &-1.841359 & -1.612913 & 0.000000 & -1.417988\\
10& -1.431632 & -1.539663 & 0.000000 & -0.235633\\
\hline
\end{tabular*}
\end{table}

Finally, we see how our model performs versus other published methods
when searching for transcription factor binding sites along the genome.
In Section~\ref{secExample} we observe that for the transcription
factor Bicoid there is good agreement between the putative binding
sites predicted by an in-vivo ChIP-chip experiment performed in the
Biggin Lab and all the other published methods we compare it with;
however, the agreement is strongest between the ChIP-chip experiment
and the results of our model applied to the SELEX data for Bicoid.

We have chosen to explain the results from Bicoid in detail because it
has been studied extensively in the literature and we have
multiple replicates of the SELEX experiment, the multiplex assay
experiment and the ChIP-chip experiment. The protocol for the SELEX
experiment is provided in \citet{Ogawa}.

\subsection{Simulations}\label{secSim}

To explore the properties of our estimation procedure, we simulated
data under our model and refit the model parameters from the
simulated data. The energy model that we simulated under is a plausible
model for binding of the Bicoid homebox which is strongly attracted to
sequences that include {\texttt{TAAT}}. In fact, the energy matrix used for
the simulation is the matrix estimated in Table~\ref{tabmatrix}.

To simulate data under the SELEX model, we generated one million 16mer
random sequences uniformly, which we refer to as round 0.
Then, for rounds $r=1,\ldots,4$, we keep each
sequence in round $r-1$ with the probability given by (\ref{eqnprob}).

To simulate the PCR duplication process, in which the number of
oligonucleotides is typically much larger than the number of PCR
molecules, we repeatedly selected a sequence at random and duplicated
the selected sequence, until we had one million sequences.

After
we had reached one million
sequences, we randomly sampled $2000$ of these without
replacement. The
$2000$ sequences are the data for round $r$, which we fed into our
model. The
other sequences formed the selection pool for round $r-1$.

\begin{figure}

\includegraphics{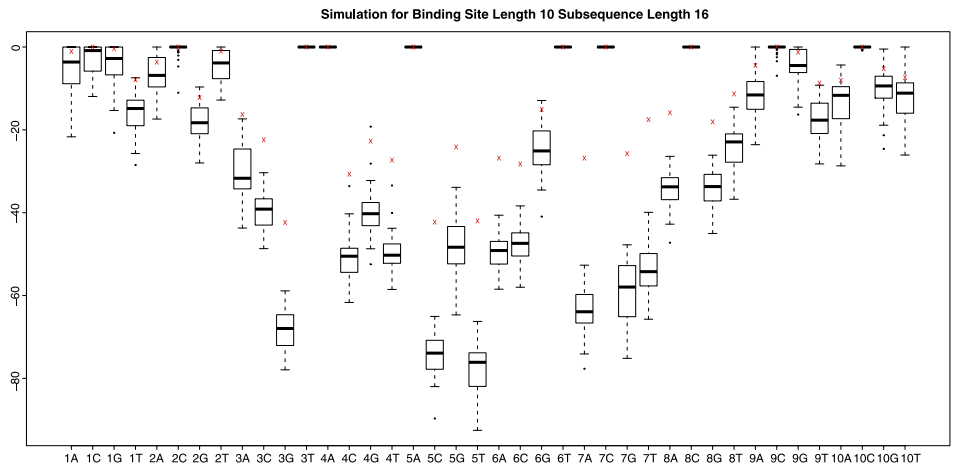}

\caption{Boxplots of the free energy parameters estimated from the
$32$ simulations. The values that generated the simulations are shown
by red crosses.}\label{figenergy}
\vspace*{-2pt}
\end{figure}

In Figure~\ref{figenergy} we present boxplots of the estimated
parameter values for $32$
simulations. We simulate our Bicoid SELEX data situation of having a
binding site length\vadjust{\goodbreak} of $l=10$ inside random $16$mer sequences $S_i$. As
can be seen, under the model,
our procedure provides biased results. In our simulations, the binding
strength of the consensus
sequence is overestimated. We believe this bias will also be present
but hopefully smaller in magnitude when real SELEX data
is analyzed, since a real SELEX experiment will begin in round 0 with
many, many more sequences that 1~million. To the best of our knowledge,
the bias is present due to the fact that we assume in our model that
every sequence type $S_i$ is present in each round $r$ of the SELEX
experiment. In reality, of course, and in our simulations, weaker
sequences will not make it to later rounds of SELEX. This will make the
consensus sequence look stronger than it really is.

Many more simulations are provided in the supplementary material [\citet{Atherton12}]. It
appears that as the stringency of the experiment is increased (either
by decreasing the amount of the transcription factor or by increasing
the energy matrix) the bias is increased. Also seen in our simulations
in the supplementary material [\citet{Atherton12}] is that the bias is also present and much
bigger in magnitude in the BEEML model. Of course, the BEEML model
makes the same assumption as us that every sequence type is present in
each round of SELEX.

\begin{figure}[b]

\includegraphics{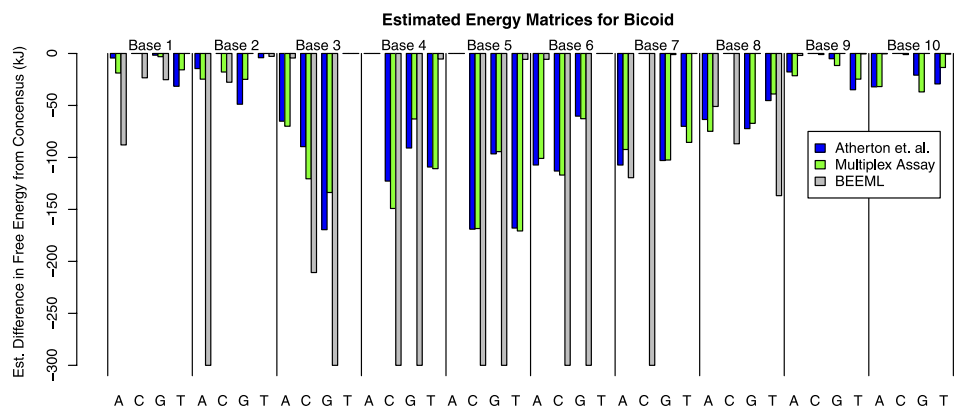}

\caption{Estimated $\Delta\Delta G$ matrices from (1) a multiplex
assay experiment from the Biggin Lab (green), (2) our model applied to
all four rounds of a SELEX experiment for Bicoid (blue), and (3)~the
BEEML model of Zhoa, Granas and Stormo (\protect\citeyear{StormoZ}) applied
to data from rounds three and four of the same SELEX experiment for Bicoid (grey).}
\label{figmultiplexassay}
\end{figure}

Unfortunately it is impossible to know exactly what sequence types are
in each round $r$ of a SELEX experiment. Since we wanted to include
data from all rounds of a SELEX experiment and include alignment in our
model, we are forced to assume that all possible sequence types are
present in every round. In our earliest efforts to model SELEX data we
used models which only included the last round of SELEX and we assumed
that the only binding sites $b_i$ that were present were the binding
site types that were observed. Of course these models required a
pre-alignment step and could only accept data from the last round of
SELEX.

\subsection{Bicoid SELEX data}\label{secStormo}

Our SELEX model was run on output from all four rounds of the Bicoid
SELEX experiment. Here $k=16$ and $l=10$. The $\Delta\Delta G$ matrix
is given in Table~\ref{tabmatrix}. The sequence with the highest
affinity to Bicoid is called the consensus sequence. Our consensus
sequence is {\texttt{GGATTAGGGG}} (or equivalently {\texttt{CCTAATCCCC}}). We
have set the energy of the consensus sequence to be $0$ in Table \ref
{tabmatrix}.

\subsection{Comparison to the multiplex assay experiment and
BEEML}\label{secmultiassay}

In addition to SELEX, the Biggin Lab has also produced an in-vitro
multiplex assay experiment. In this multiplex assay experiment a small
number of sequence types $S_i$ are produced. Usually the consensus
sequence is known a~priori (e.g., from a~SELEX experiment) and
the sequence types $S_i$ produced for the experiment vary from the
consensus sequence at one or two positions only. As in the SELEX
experiment, the $S_i$ are entered in solution with the transcription
factor Bicoid. The solution is allowed to reach equilibrium and then
the bound sequences are separated from the transcription factor. Since
there are very few sequence types $S_i$ present in this experiment, one
can obtain a much more accurate measure of the amount of bound~$S_i$
than in a SELEX experiment. Using the thermodynamic concepts presented
in this paper, one can easily use the measured amounts of each bound
sequence type to directly calculate a $\Delta\Delta G$ matrix. The
results of the multiplex assay experiment described above for Bicoid
are shown in green in Figure~\ref{figmultiplexassay}.

To compare our model to the BEEML model of \citet{StormoZ}, we had to
pre-align the sequences $S_i$ for a binding site of length ten. To do
the alignment, we used MEME [\citet{MEME}]. We considered using MEME to
directly align the sequences and then input these sequences into the
BEEML model; however, when aligning sequences MEME clusters like
sequences together and also eliminates sequences which do not fit
according to their model. Hence, we decided it was preferable to run
MEME and construct a mean PWM based on the output from round four of
the SELEX experiment. We then used the PWM to find the highest affinity
subsequence of length ten in each 16mer $S_i$ from rounds three and
four of the SELEX experiment. These subsequences were the aligned
binding sites that were given to the BEEML model as input. The results
of BEEML are shown in grey in Figure~\ref{figmultiplexassay}.

Finally, as described in Section~\ref{secStormo}, we ran our model on
all rounds of a~SELEX experiment for Bicoid. The results are plotted in
Figure~\ref{figmultiplexassay} in blue.

From Figure~\ref{figmultiplexassay} we see that the consensus sequence
for the multiplex assay experiment is {\texttt{CTTAATCCCC}} and the
consensus sequence for BEEML is {\texttt{TGTAATTGGG}}. Recall from Section
\ref{secStormo} that the consensus sequence for our model is {\texttt{CCTAATC}}
\texttt{CCC}. It is clear that all three models pick up the {\texttt
{TAAT}} homebox which is clearly the most important factor in
determining the affinity of a subsequence to Bicoid. Also seen from
Figure~\ref{figmultiplexassay} is how deleterious a mutation in the
homebox is to binding. Any mutation from {\texttt{TAAT}} at positions three
to six leads to a very substantial decrease in $\Delta\Delta G$. All
models show that mutations from the consensus sequence at positions
nine and ten are not very critical to binding. The three models also
indicate that positions one and two are weakly critical to binding;
however, BEEML indicates it is deleterious to have nucleotide base {\texttt
{A}} at positions one and two, whereas our model and the multiplex
assay do not show the same deleterious effect. There are other obvious
instances where the BEEML model deviates significantly from our model
and the multiplex assay experiment.

As for why the BEEML model deviates quite a bit from our model and the
multiplex assay experiment for certain nucleotide estimates at certain
locations, a~main reason is most likely the need to pre-align using
MEME, that is, the output we see for BEEML will be heavily influenced
by MEME. Our model aligns during the optimization of the likelihood
and, hence, unlike MEME, our alignment is based on thermodynamic
principles. There are also important differences between BEEML and our
model. Both BEEML and our model are thermodynamic models run on the
same SELEX experiment, however:
\begin{itemize}
\item BEEML accounts for the nonspecific energy of binding. Although
our model can account for the nonspecific binding, in this instance, it
was run without accounting for nonspecific binding.
\item BEEML accounts for errors in the PCR step. We have chosen not to
account for that explicitly in our model.
\item BEEML also has an expression similar to our expression (\ref
{eqnnojunkprobM}) for $P_r(S_i)$. The problem both models encounter is
that there are too many terms to enumerate in the denominator. As
described in Section~\ref{secden}, we use Monte Carlo to overcome
this. The BEEML model takes a different approach similar to \citet
{Djordjevic2006} where they discretize over a~user defined number of
energy levels.
\item Our model uses data from all rounds of the experiment.
Furthermore, we carefully model the sequence enrichment from one round
to the next. The code for BEEML accepts data from two rounds of SELEX,
however, there is no indication in \citet{StormoZ} that they correctly
model the progression from one round to the next.
\item The final likelihoods for our model and BEEML are different and
optimization schemes used are also different.
\end{itemize}

Hence, although the BEEML model has offered significant improvements to
the original \citet{Djordjevic2006} model, we believe that our model
offers further important improvements.

Of course, we also see that our model estimates deviate slightly from
the multiplex assay estimates and we hope that in these instances our
model is providing good estimates for the $\Delta\Delta G$ matrix
since we are using data from many, many more sequence types $S_i$ than
the multiplex assay experiment. In particular, we are including
sequences with a full range of affinities from low to high.

As the $\Delta\Delta G$ energies from the multiplex assay are
calculated directly from the thermodynamic equations, we do not
anticipate a big bias in the multiplex assay estimates. There does not
seem to be any consistent difference between our model estimates of
$\Delta\Delta G$ and the estimates from the multiplex assay
experiment. This observation supports our hypothesis that the bias
observed in our SELEX simulations will be reduced when our model is
applied to real data since in a real SELEX experiment there are many
more sequences present and, hence, many more low affinity sequences
will make it through to later rounds than in our simulation. Basically,
we think that the assumption of each sequence type being present in
each round is more valid in the real data situation than in the
simulated data situation.
\begin{figure}

\includegraphics{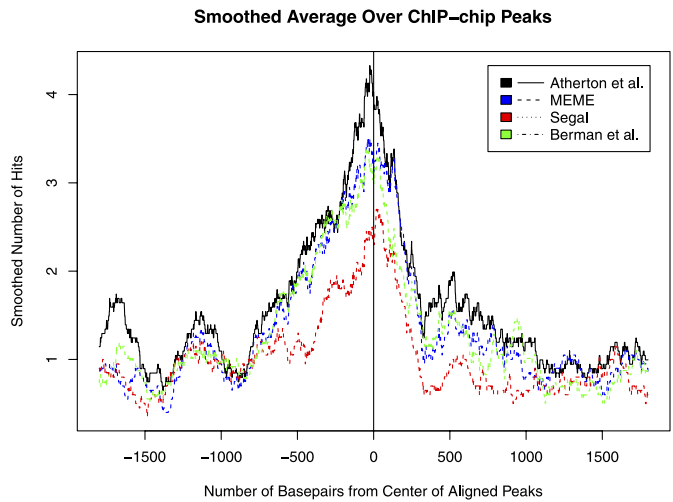}

\caption{Smoothed average of predicted binding sites for four models
at ChIP-chip peaks. The legend is as follows: Atherton et al.
represents the model discussed in this paper, MEME represents
Bailey et~al. (\protect\citeyear{MEME}), Segal represents Segal et~al.
(\protect\citeyear{SegalandSadka}) and Berman et al.
represents Berman et al. (\protect\citeyear{Berman}). The fixed parameters
(as described in
Appendix \protect\ref{appChip}) for the analysis of the ChIP-chip data are
$n_p = 100$, $w_s = 4000$, $n_s = 100$, and $s_t = 0.999$. The peaks
are aligned so that the center of each peak, defined as the highest
point in the peak, appears at $0$ on the $x$-axis.}\vspace*{-3pt}
\label{figchipcomparison}
\end{figure}

\subsection{Comparison in an in-vivo setting}\label{secExample}

Using the $\Delta\Delta G$ matrix estimated by our model on the Bicoid
SELEX data in Table~\ref{tabmatrix},
we scan the genome of Drosophila Melanogaster and compare the results
of our model and three other popular models to the results of an
in-vivo ChIP-chip experiment.

The Berkeley Drosophila Transcription Network Project (BDTNP) has
generated SELEX and ChIP-chip data for Bicoid. ChIP-chip data
measures the genome wide relative levels
of occupancy for a single protein of interest. We used the BDNTP
ChIP-chip data and a simple, nonparametric method to validate and
compare our Bicoid model with a PWM derived from MEME [\citet{MEME}] and two
models from the literature [\citet{SegalandSadka} and \citet{Berman}].
All four methods show strong agreement with the in-vivo ChIP-chip data,
however, our
model has the strongest agreement; see Figure~\ref{figchipcomparison}.

The ChIP-chip experiments identified thousands of genomic
regions to which Bicoid binds. This data has been shown to provide a
quantitative measure of relative occupancy. That is, regions can be
assigned a score, and those scores have been shown to be reproducible
between biological replicates [\citet{Li} and \citet{MacArthur}]. From these
and other observations, the authors concluded that the high scoring
regions correspond to those with the highest net occupancy of bound
factor.

Because of the complexity of intracellular processes, a binding model
alone does not provide enough information to predict the results of a
ChIP-chip experiment. For instance, without additional data, we have
no way of modeling the inhibitory affect of chromatin structure.
However, we can still use the identified binding regions to test the
validity of our SELEX model and data.

If a binding model is identifying true in-vivo binding sites, then we
expect the number of high affinity sites predicted by our model to be
higher near ChIP-chip peaks. Roughly, we compared the binding models
by measuring the enrichment of identified binding sites as compared to
the genomic background. There were several variables that we
controlled for; we explain the method in detail in Appendix~\ref
{appChip}. We
plotted the results of this analysis for our model and competing models
in Figure~\ref{figchipcomparison}.

%

Absent from our comparison in Figure~\ref{figchipcomparison} is the
\citet{StormoZ} model. Since, as discussed in Section \ref
{secmultiassay}, we have to
pre-align the sequences of the SELEX experiment using MEME, the output
in Figure~\ref{figchipcomparison} after transformation by the
sequence ranks will be
very near to the output of MEME presented in
Figure~\ref{figchipcomparison}.


\section{Conclusion}\label{secCon}
The model presented here attempts to infer a comprehensive map of the
sequence specific binding affinities between double stranded DNA and a
transcription factor from a~SELEX experiment. There exist a~variety of
assays, including ChIP-chip, that attempt to measure the average
binding behavior of a protein in a population of cells. However, only
in vitro assays like SELEX can provide precise thermodynamic models of
protein/DNA interactions for downstream models of transcriptional control.

To make accurate inference from SELEX data, researchers have left the
traditional empirical approaches such as PWMs and recently turned to
creating models for SELEX based on the physical chemistry of binding.
The goal of these models is to estimate the free energy of binding,
$\Delta G$, matrix.
Often the exact binding site length $l$ is unknown a priori, hence,
SELEX experiments are performed with a sequence length $k$ greater than
$l$. Also, by taking a large $k$, as in the Biggin Lab, once a random
pool of sequences has been generated, SELEX experiments can be
performed for many transcription factors with varying binding site
lengths $l$. Our model for SELEX is the first model capable of
accepting data of the form $k > l$. Other models for SELEX can only
accept data with $k=l$ or require an alignment step a priori. Another
important feature of our model is that it accepts data from all rounds
of the SELEX experiment. This is crucial for estimation of $\Delta G$,
since a mix of oligonucleotides that have a range of affinities for the
transcription factor are required. Previous models only use data from
the last round of the SELEX experiment and hence base their estimates
on oligonucleotides with a high affinity to the transcription factor.

The success of our model is demonstrated by applying our model and
three others to predict the DNA recognition sites enriched in an
in-vivo ChIP-chip experiment. The in-vivo ChIP-chip experiment
indicates the in-vivo occupancy of the transcription factor along the
genome. A prior, it may not have been the case that the affinity of a
sequence for a transcription factor as measured in an in-vitro
experiment is a good predictor for binding sites occupied in-vivo, even
after taking into account of the influence of other proteins, such as
nucleosomes, on occupancy in vivo. However, we have found that for the
transcription factor Bicoid the recognition sites used in-vitro and
in-vivo are very closely related. Hence, we can use the in-vivo
ChIP-chip experiment as validation when comparing different models and
motifs for binding. It is important that a comparison of models be made
with the ChIP-chip experiment, as this can serve as a gold standard for
binding affinity; otherwise, finding that two models produce different
motifs or different energy matrices is insufficient to determine which
model is performing better. Our success using results from an in-vivo
experiment to validate the results of an in-vitro experiment suggests
that SELEX does provide a quite accurate, fine scale model of the
intrinsic DNA recognition properties of a transcription factor. The
results of our comparison in Section~\ref{secExample} demonstrate that
our model outperforms the other models.

Preliminary results suggest that varying the additive $\Delta G$
parametrization of our model would provide the biggest predictive
improvement. For instance, base pair dependencies can be added.
Alternatively, one could take a feature based approach; see \citet
{Segal}. In the case of Bicoid, a feature based approach could
specifically model the {\texttt{TAAT}} homebox.

\begin{appendix}\label{app}
\section{Chemical concepts}\label{secchem}

The concepts introduced here can be found in the physical chemistry
textbook by \citet{Atkins}.
We begin by considering many copies of a single oligonucleotide species
$S$ in solution with a transcription factor $\mathit{TF}$.
Furthermore, we assume that~$S$ and $\mathit{TF}$ always bind in the same configuration.

When $S$ and $\mathit{TF}$ are entered into solution with one another they will
react to
form the product $\mathit{TF}\dvtx  S$. We call this the {\textit{forward
reaction}}. The
product $\mathit{TF}\dvtx  S$ will also disassociate into $S$ and $\mathit{TF}$; we call
this the
{\textit{backward reaction}}. The following chemical equation,
\[ 
\mathit{TF} + S \rightleftharpoons \mathit{TF}\dvtx S,
\]
represents these reactions. The solution is said to be in
{\textit{dynamic equilibrium}} when the forward rate of reaction equals the
backward rate of reaction. A~dimensionless physical constant
quantifying the
dynamic equilibrium is the {\textit{equilibrium constant}}~$K$. Our
interest in $K$
is that it relates directly to the change in Gibbs free energy,
$\Delta G$, for the reaction. The change in Gibbs free energy, $\Delta
G$, quantifies
the affinity of $S$ for $\mathit{TF}$. Hence, in
Section~\ref{secPS} we parameterize our SELEX model in terms of
$\Delta G$.

Letting $R_{\mathrm{Gas}}$ represent the ideal gas
constant and $T$ the temperature in Kelvins, we have
\begin{equation}\label{eqnrelation}
K = \exp\biggl(-\frac{\Delta G}{R_{\mathrm{Gas}}T}\biggr).
\end{equation}

As we shall see below, $K$ is unidentifiable without meta data. The
meta data was defined in Section~\ref{secSELEX}.

The forward rate of reaction is proportional to the product of
concentrations of
the reactants. The {\textit{forward rate constant}}, $k_f$, is the proportionality
constant. Hence,
%
\begin{equation}\label{eqnfor}
\mathrm{Forward\ rate} = k_f[S][\mathit{TF}]
\end{equation}
and, similarly,
%
\begin{equation}\label{eqnbac}
\mathrm{Backward\ rate} = k_b[\mathit{TF} \dvtx  S].
\end{equation}
At equilibrium, equating (\ref{eqnfor}) and (\ref{eqnbac}) gives
the following expression for the equilibrium constant $K$:
%
\begin{equation}\label{eqnK}
K = \frac{k_f}{k_b} = \frac{[\mathit{TF}\dvtx S]}{[\mathit{TF}][S]}.
\end{equation}

We can think of $K$ as an expected value where the ``concentrations''
are averages
over time and space. In principle, we can use
the {\textit{observable}} concentrations $\widehat{[S]}$, $\widehat{[\mathit{TF}]}$ and
$\widehat{[\mathit{TF}\dvtx S]}$ to estimate the theoretical physical
quantity~$K$ and in turn $\Delta G$ [via (\ref{eqnrelation})].

\section{Identifiability}\label{appidentifiability}

There are three types of lack of identifiability in the SELEX model
outlined below.

\subsection{\texorpdfstring{Identifiability between $[\mathit{TF}]_r$ and $\Delta G$}{Identifiability between $[\mathit{TF}]_r$ and Delta G}}\label
{appdeltadeltaG}

The structure of $\widehat{t_r}(S_i)$ in~(\ref{eqntau}) reveals that
the $\Delta G(b_j)$s are not directly identifiable without knowledge of
$\widehat{[\mathit{TF}]_r}$. This is because $\widehat{t_r}(S_i)$ is unchanged
by rescaling all the $\Delta G(b_j)$s and $\widehat{[\mathit{TF}]_r}$
by the same constant. However, with the given data, we can always estimate
\[
\Delta\Delta G(b_j) = \Delta G(b_j) - \Delta G(b_o),
\]
where $b_o$ is a reference binding site such as a consensus sequence.
Of course, if we have meta data such as $\widehat{[\mathit{TF}]_r}$, we can
estimate $\Delta G(b_j)$.

\subsection{\texorpdfstring{Identifiability in additive $\Delta G$}{Identifiability in additive Delta G}}
Physically, we are able to identify the total binding affinity of a
binding configuration but not the contributions of the individual base
pairs. To solve this, we choose to fix the energy of the highest
affinity base pair in each position except one to be zero. Then, the
value of the first position's highest energy base pair is interpretable
as the binding affinity of the ``consensus sequence,'' or the modeled
highest affinity binding site. Some care is needed in ensuring that
this constraint does not interfere with whatever optimization algorithm
is chosen---such concerns are discussed in the code's comments.

\subsection{Identifiability of the binding site names} \label{appIden3}
The third identifiability problem is present in any binding model which
represents binding sites by their sequences. For any segment $b_j$ of a
double stranded DNA sequence there are four possible names. To ensure
that the paramterization is physically meaningful, each binding site
must be represented by the same sequence. For example, Bicoid has a
high affinity for sequences that contain the subsequence {\texttt{TAATCC}}.
As can be seen in Table~\ref{tabSeqAlign}, it is possible to align the
full sequences by the subsequences that are closest to {\texttt{TAATCC}} in
the Hamming sense. If, for instance, one were to name half of the
subsequences by {\texttt{TAATCC}} and half by {\texttt{ATTAGG}}, then the
likelihood would not optimize properly. This being said, it is
irrelevant which name is chosen, as long as it is consistent. For
instance, the subsequence {\texttt{TAATCC}} could also be called {\texttt
{CCTAAT}}, {\texttt{ATTAGG}} or {\texttt{GGATTA}}. For the binding model
presented in Section~\ref{secbinding}, the likelihood will be
symmetric with four identical modes, each corresponding to a~different
naming scheme for the strongest binding site. Which of the names our
code chooses is chosen, arbitrarily, to be the one with the consensus
sequence, that is, first alphabetically.

\begin{table}
\tablewidth=200pt
\caption{Possible binding sites of length $l=10$ for the factor Bicoid
in an oligonucleotide of length 16}\label{tabalignseq}
\begin{tabular*}{200pt}{@{\extracolsep{\fill}}lcc@{}}
\hline
\multicolumn{1}{@{}l}{\textbf{3}$\bolds{'}$} & \textbf{{\texttt{GTTTATAATCCGCGTC}}} & \multicolumn{1}{c@{}}{\textbf{5}$\bolds{'}$} \\
\hline
& {\texttt{CAAATATTAGGCGCAG}} & \\
1 & {\texttt{GTTTATAATC\hspace*{0.446 in}}}& \\
2 & {\texttt{\hspace*{0.074333 in}TTTATAATCC\hspace*{0.38566 in}}} & \\
3 & {\texttt{\hspace*{0.14866 in}TTATAATCCG\hspace*{0.31133 in}}} & \\
4 & {\texttt{\hspace*{0.22299 in}TTATAATCCG\hspace*{0.237001 in}}} & \\
5 & {\texttt{\hspace*{0.297332 in}TATAATCCGC\hspace*{0.162668 in}}} & \\
6 & {\texttt{\hspace*{0.371665 in}TATAATCCGC\hspace*{0.088335 in}}} & \\
7 & {\texttt{\hspace*{ 0.445998 in}TATAATCCGC\hspace*{0.014002 in}}} & \\
\hline
\end{tabular*}
\end{table}
%

\section{Description of ChIP-chip comparison}\label{appChip}
We compare the predictions for putative binding sites for Bicoid from
our SELEX model and experiment to predictions from \citet{MEME}, \citet
{SegalandSadka} and \citet{Berman}. For validation, all four models,
ours, \citet{MEME}, \citet{SegalandSadka} and \citet{Berman}, are used to
predict the putative binding sites at genomic locations previously
highlighted in a ChIP-chip experiment. In \citet{MacArthur} they defined
a ``peak'' of the ChIP-chip experiment to be a single point in the
genome where the local signal achieves its maximum. In our
nonparametric comparison of the models for the binding affinity of
Bicoid we chose to consider the $n_p$ highest peaks in the ChIP-chip
experiment. To summarize our results, for each of the four models we
combine the putative binding site predictions over the $n_p$ peaks in
the method described below. Note that since some models attempt to
assign physically meaningful affinity scores to each subsequence (e.g.,
the use of the free energy matrix in our model) and other models assign
affinity scores based on estimated probabilities or background
frequencies (e.g., the use of the PWM in MEME), an important step of
our comparison is to obtain a common scoring scale for the four models.
In Section~\ref{seccompare} we explain how we obtain the common
scoring scale. For each of the four models, the steps in Section~\ref
{seccompare} are repeated at each of the ChIP-chip peaks. Section~\ref
{seccombine} explains how we combine and summarize the results of the
$n_p$ peaks for each model.

\subsection{Common scoring scale}\label{seccompare}
To obtain a common scoring scale for the four models, for each model it
is necessary to relate the affinity scores at the peaks to the affinity
scores in the noncoding genome. Therefore, for each model we begin by
sampling $n_s$ intervals of size $2w_s$ from the noncoding mappable
genome that do not overlap regions identified by the ChIP-chip
experiment. Within each of the~$n_s$ intervals, we evaluate the
affinity score of each subsequence of length $l$, thus generating $n_s$
samples of affinity scores. Each sample provides an empirical null
distribution of affinity scores. We choose an $\alpha$ (e.g., $\alpha
=0.01$), and in each of the $n_s$ samples we find the $\alpha
$th-percentile affinity score. To calculate a threshold affinity score,
we take the median of the $n_s$ $\alpha$th-percentiles. Our threshold
affinity score is denoted by~$\widehat{s}_\alpha$.

Next, for each model, we examined a symmetric interval of fixed size~$2w_s$
around each ChIP-chip peak.
Within each of these intervals, using the chosen model, we evaluated
the affinity score of each subsequence of length~$l$.
For each subsequence of length~$l$ in the~$2w_s$ interval around each
of the~$n_p$ peaks, we consider a position to be a ``hit'' if its score
is greater than~$\widehat{s}_\alpha$.

In this way, by determining if each sequence of length $l$ near each
ChIP-chip peak is a hit or not, we can compare the four models.

\subsection{Combining the results for the $n_p$ peaks}\label{seccombine}
For each model and each peak, by defining each hit as a 1 and each
``miss'' as a 0, we obtain a binary vector that records each position
at which a hit begins. For each model, we align the $n_p$ vectors at
the peaks in the 5$'$--3$'$ direction and sum across them. The resulting
vector of counts records, with respect to the position of peaks, how
many of the $n_p$ intervals had a hit at each relative position. We
smooth these counts with a 200bp moving average,\setcounter{footnote}{4}\footnote{The 200bp is
motivated by the fact that in the ChIP-chip assay proteins bind to DNA
fragments of roughly 200 bps.} and then divide the result by the
expected number of hits under a uniform null, $n_p (1 - \widehat
{s_{\alpha}})^{- 1}$. It is these smoothed results that are plotted for
each of the four models in Figure~\ref{figchipcomparison}.
\end{appendix}

\section*{Acknowledgments}
Thanks to John Atherton,
Stephanie Atherton and Alex Glazer for helpful discussions regarding
physical chemistry.

\begin{supplement}[id=suppA]
\stitle{Code for SELEX model}
\slink[doi]{10.1214/12-AOAS537SUPP} 
\slink[url]{http://lib.stat.cmu.edu/aoas/537/supplement.pdf}
\sdatatype{.pdf}
\sdescription{The code for the SELEX model used in the application of
this paper is available at the above url. Extra simulations, mentioned
in Section~\ref{secSim}, are also provided as supplementary material.}
\end{supplement}

%

%

\printaddresses

\end{document}